\documentclass[aps, reprint, amsmath, floatfix]{revtex4-1}
\usepackage{amsthm}
\usepackage{mathtools}
\usepackage{xcolor}
\usepackage{graphicx}
\usepackage[left=23mm,right=13mm,top=35mm,columnsep=15pt]{geometry} 
\usepackage{adjustbox}
\usepackage{placeins}
\usepackage[T1]{fontenc}
\usepackage{lipsum}
\usepackage{csquotes}
\usepackage{bm}
\usepackage{comment}
\usepackage[normalem]{ulem}
\usepackage[subrefformat=parens]{subcaption}
\begin{document}
\title{The kernel-balanced equation for deep neural networks}

\author{Kenichi Nakazato}
    \email[Correspondence email address: ]{kenichi_nakazato@sensetime.jp}
    \affiliation{SenseTime Japan Ltd., Minato-ku, Tokyo, Japan}

\date{\today} 

\begin{abstract}
Deep neural networks have shown many fruitful applications in this decade. A network can get the  generalized function through training with a finite dataset. The degree of generalization is a realization of the proximity scale in the data space. Specifically, the scale is not clear if the dataset is complicated. Here we consider a network for the distribution estimation of the dataset. We show the estimation is unstable and the instability depends on the data density and training duration. We derive the kernel-balanced equation, which gives a short phenomenological description of the solution. The equation tells us the reason for the instability and the mechanism of the scale.
The network outputs a local average of the dataset as a prediction and the scale of averaging is determined along the equation. The scale gradually decreases along training and finally results in instability in our case.
\end{abstract}

\keywords{deep neural networks, training response, uncertainty estimation}

\maketitle

\section{Introduction}\label{sec:introduction}
In the recent decade, data-driven modeling has been empowered with techniques from machine learning. Among them, deep neural networks are the most powerful ones with a large number of applications\cite{AlphaGo,med_image,med_vision,chatgpt4,auto_driving}. Despite the fruitful ones, we do not know much about the mechanism behind them\cite{deepL,deepL2}. Specifically, the network can get generalized functions only with the finite dataset. In other words, the network can learn a generalized relation between input and output from the finite one. We can get predictions for unknown inputs but do not fully understand how it works.

A neural network, $y=f(\bm{x},\bm{w})$, can be defined with an input, $\bm{x}$, and output, $y$. In training, we adjust the parameters, $\bm{w}$, of the network so that the pre-defined relation, $y_i=f(\bm{x}_i)$, is satisfied as possible. As the pre-defined relation, we give a dataset, $\{(\bm{x}_i,y_i)\}$, in advance. We usually update the parameters step by step along the gradient, $\sum_i\nabla L(|y_i-f(\bm{x}_i)|)$, of the minimizing function, $L$. The name \textit{neural network} stems from the architectures of the function, $f$, which is inspired by the brain network. In fact, one of the most famous architecture, convolutional neural networks, is originally a model of retinal structure\cite{cnn1,cnn2,cnn3,cnn4}. In this paper, we focus on them.

We can derive neural tangent kernels, NTKs, or training responses as a theoretical approach to understanding the generalization mechanism\cite{NTK1,NTK2,NTK3,NTK4,TR}. There, we can describe the training response, $\Theta(\bm{x},\bm{x}_i)$, which shows the influence on the output, $f(\bm{x})$, by a training step, in the following equation,
\begin{eqnarray}
f(\bm{x},\bm{w}+\bm{\delta})&\sim& f+\frac{\partial f}{\partial \bm{w}}\cdot\bm{\delta}\\
&=&f-\eta\frac{\partial f}{\partial \bm{w}}\cdot\frac{\partial f_i}{\partial \bm{w}}\frac{dL}{df_i}\\
&\equiv&f-\eta\Theta(\bm{x},\bm{x}_i)\frac{dL}{df_i},
\label{eq:NTK}
\end{eqnarray}
where the model is trained with a single data, $(\bm{x}_i,y_i)$, with a minimizing target, $L$, known as loss function and the parameter, $\eta$, is a learning rate. We usually have a dataset with many data points, $\{(\bm{x}_i,y_i)\}$, and train the model with that. In such a case we can write the equation with the sum of training responses,
\begin{equation}
\Delta f(\bm{x})\propto -\sum_j\Theta(\bm{x},\bm{x}_j)\frac{dL}{df_j}.
\label{eq:NTKs}
\end{equation}
Furthermore, in some cases, we can assume a simple ansatz for the training response with an aging effect. It can be expressed in the following, 
\begin{equation}
\Theta(\bm{x},\bm{x}_i)\propto t^{-\alpha}K(\bm{x},\bm{x}_i),
\label{eq:ansatz}
\end{equation}
where the exponent, $\alpha$, shows aging decay and the response kernel, $K$, is a positive decreasing function of the distance, $|\bm{x}-\bm{x}_i|$. The decreasing scale would be determined by the architecture, but we assume it is a similar one with an exponential curve, in this paper.

As a minimizing function, we consider it a more complicated problem than simple supervised training. In standard supervised training, we optimize a network so that the relation, $y_i=f(\bm{x}_i)$, is satisfied for any data point. On the contrary, we want to estimate the distribution of the dataset, $\{(\bm{x}_i,y_i)\}$. To do that, we estimate the local mean, $\mu(\bm{x})$, and standard deviation, $\sigma(\bm{x})$, for each input, $\bm{x}$. As a network, we assume the following one,
\begin{eqnarray}
\mu(\bm{x})&=g\circ f(\bm{x})\label{eq:gf}\\
\sigma(\bm{x})&=h\circ f(\bm{x}).\label{eq:hf}
\end{eqnarray}
We have a shared function, $f$, and specific ones, $g$ and $h$, to estimate the mean and standard deviation, respectively.
As an optimizing function, we minimize the following,
\begin{equation}
L\equiv -\log(\Pi_i\frac{1}{\sigma(\bm{x}_i)\sqrt{2\pi}}\exp(-\frac{1}{2}(\frac{y_i-\mu(\bm{x}_i)}{\sigma(\bm{x}_i)})^2)).\label{eq:Gloss}
\end{equation}
In other words, we want to fit with a Gaussian distribution. This type of problem is known as \textit{uncertainty estimation} in the field of machine learning\cite{uncertainty_rev,Bdeep,uncertainties}. We want to know both maximum likelihood estimation, $\mu(\bm{x})$, and uncertainty of that, $\sigma(\bm{x})$, at the same time. Our problem setting is a much simpler one than various applications.

In the case of the standard prediction, we usually minimize the distance, $|\mu_i-y_i|$, and the optimal solution is the exact one, $\mu_i=y_i$. However, we can have a different mean value because we simultaneously estimate standard deviation, $\sigma$, in our network. As stated above, training with a data, $(\bm{x}_i,y_i)$, can influence on another prediction for a data point, $(\bm{x}_j,y_j)$. In sum, the predicted mean and standard deviation, $\mu(\bm{x}_i)$ and $\sigma(\bm{x}_i)$, can be a local estimation of the data distribution. Our main question is how the scale of estimation is determined. {\color{black} We assume two different inputs, $\bm{x}_a$, and $\bm{x}_b$, should have similar outputs, $f(\bm{x}_a)\sim f(\bm{x}_b)$, depending on a distance between them, as the nature of prediction. However, we do not know the reality in the cases of deep neural networks.} In other words, we want to know how and to what extent the estimation is generalized.

{\color{black} As a hypothesis, we can assume some specific scales of the structure of the dataset are reflected in the prediction. In other words, the scale may be the reflection of the semantic structure of the dataset. However, we use a randomly generated dataset rather than specific public ones, like the other studies in the statistical physics\cite{sopt,copt}. The advantage of this way is that we can get a universal understanding of the nature of the neural networks independent of the dataset instance.}

In the next section \ref{sec:model}, we introduce our model. There we describe our network and dataset. In addition, we note on training method as well. We show the training dynamics in the section \ref{sec:results}. We show the estimation is unstable. Furthermore, we introduce a phenomenological description, kernel-balanced equation, of the solution, which explains the instability. It gives answers, the scale and generalization, for our question. Finally, we show that the equation can be redisplayed with dynamics by training response.
\section{model} \label{sec:model}

In general, deep neural networks consist of layers of nonlinear transformations, $f_i$, and an input, $\bm{x}$, and output, $y$,
\begin{equation}
y=f_n\circ\cdots\circ f_0(\bm{x}).
\end{equation}
In each layer, we often use combination of linear convolution, $c_{jk}$, and non-linear activation function, $R$,
\begin{equation}
\bm{h}_{k}=R(b_{k}+\sum_jc_{jk}(\bm{h}_{j})),
\end{equation}
where j-th channel of input for a layer, $\bm{h}_j$, is transformed into k-th channel of output, $\bm{h}_{k}$. Hidden variables, $\bm{h}_i$, have often multi-channels for more degree of freedom of the network. We call this as a convolution layer\cite{cnn1,cnn2,cnn3,cnn4}.


Here, we focus on a simple convolutional network with $n$ layers. We assume the input, $\bm{x}$, is a 1-dimensional bit string with the size, $b$. {\color{black} In other words, we assume the input, $\bm{x}$, is a binary vector in this paper.} Each convolution layer can be defined with the number of output channels, $s_c$, and kernel size, $s_k$, of the convolution. In our model, the number of channels, $s_c$, in each convolution layer is fixed. Needless to say, the output of a mid-layer means the input to the next layer. In the final layer, we usually use a linear network,
\begin{equation}
y=R(\sum_i\bm{a}_i\cdot\bm{h}_i+b),
\end{equation}
where the hidden variable, $\bm{h}_{i}$, is the input for the last layer. Non-linear activation, $R$, is applied after linear transformation with parameters, $\bm{a}_i$ and $b$.
In numerical experiments, we use a setting, $b=8$, $s_k=3$, $s_c=3$, with ELU as an activation function and SGD for the training algorithm with a learning rate, {\color{black} $\eta=0.1$}, without a learning momentum\cite{elu,sgd1,sgd2,sgd3}.

Since we consider a network with two outputs, $\mu$, and $\sigma$, we have two linear networks, $g$ and $h$, after the convolution layers, $f$, in eq. (\ref{eq:gf}) and (\ref{eq:hf}). We call the network as \textit{variance network}, here. In addition, we also consider a simpler one with only one output, $\mu$, for easier understanding, in eq. (\ref{eq:gf}). We call that as \textit{average network}.

As the dataset, we consider a random bit encoding\cite{TR}. The dataset, $\{(\bm{x}_i,y_i)\}$, consists of pairs of an input, $\bm{x}_i$, and output, $y_i$. We randomly generate $N$ pairs in advance and use them as a training dataset. We train a model with the dataset and a loss function, eq. (\ref{eq:Gloss}). We can focus on training dynamics itself independent of a specific dataset instance by testing randomly generated ones and analyzing its statistical features.


We also consider a simplified model for understanding training dynamics. As stated, training dynamics can be described simply in a simple equation, (\ref{eq:NTK}) or (\ref{eq:ansatz}). The training response, $\Theta(\bm{x},\bm{x}_i)$, is known as a neural tangent kernel and can be constant during training in an infinity limit of network size\cite{NTK1,NTK2}. Even if the size is finite, it can be represented by a product of a time-dependent term, $A(t)$, and almost constant kernel, $K(\bm{x},\bm{x}_i)$, like equation (\ref{eq:ansatz})\cite{TR}. In the case of an average network, we can write down simplified dynamics,
\begin{equation}
\frac{d\mu_i}{dt}\propto-\sum_j K(\bm{x}_i,\bm{x}_j)\frac{dL}{d\mu_j},
\label{eq:kernel_dyn}
\end{equation}
where we ignore the time-dependent term in training response. In other words, we consider short-term training dynamics and call it as \textit{response kernel dynamics}.

In many cases, the loss function, $L$, evaluates the distance between the prediction, $f_i$, and the answer, $y_i$, e.g. mean squared error. In such a case, we can write it as follows,
\begin{equation}
\frac{df_i}{dt}\propto \sum_jK_{ij}(y_j-f_j),
\label{eq:kernel_dyn2}
\end{equation}
where the response kernel, $K_{ij}$, can be seen constant during training as assumption.
\section{results} \label{sec:results}
\subsection{1-point training}
We start from the simplest one, where our dataset has only a pair, $(\bm{x}_0,y_0)$. We train a variance network with that. We call it \textit{1-point training}, here. When the dataset consists of N-pairs, we call it \textit{N-point training}.

{\color{black} Before showing results, we should confirm the form of the loss function, equation \ref{eq:Gloss}. Since the loss is the log-likelihood of the Gaussian, it can be rewritten with the summation, easily,

\begin{equation}
L\propto\Sigma_i(\sigma(\bm{x}_i)+\frac{1}{2}(\frac{y_i-\mu(\bm{x}_i)}{\sigma(\bm{x}_i)})^2).
\end{equation}

The first term, $\Sigma_i\sigma(\bm{x}_i)$, can be minimized by the minimal of the standard deviation, $\sigma(\bm{x}_i)=0$. However, the second term includes that in its denominator and can be divergent by the zero value. If the first term is minimized faster, the second term can change its value abruptly. In other words, we can see a numerical instability, there. On the contrary, if the scale of the standard deviation can be adjusted moderately, we do not see numerical instability. We can focus on the trajectory of the point, $(y_i-\mu(\bm{x}_i),\sigma(\bm{x}_i))$, to see the determination of the prediction scale.
}

\begin{figure}
\includegraphics[width=1.0\linewidth]{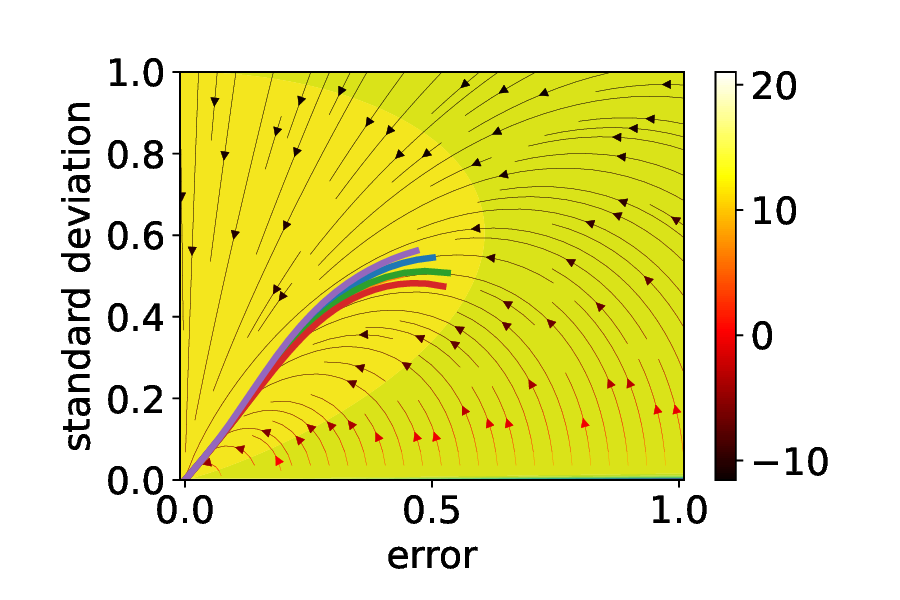}
\caption{Training dynamics with a single point data. A network is trained with a data point, $(\bm{x}_0,y_0)$, and the error, $|\mu_0-y_0|$, and standard deviation, $\sigma_0$, are shown on the map of loss function. In the figure, we show 4 results with different colors and initial conditions, but all trajectories finally end up with numerical instability. The vector field shows the gradient of the loss function. The color of the arrows shows the steepness of the gradient in the log scale. Here we used the setting, learning rate $\eta=0.1$, the size of input $b=8$, the number of layers $3$. We used SGD and ELU as an optimizer and activation, respectively.}
\label{fig:point_training}
\end{figure}

We show the results of 1-point training in FIG. \ref{fig:point_training}. In the figure, we show trajectories of training dynamics on the vector field along the loss gradient. All of them start from around the center, $|\mu_0-y_0|\sim 0.5$ and $\sigma_0\sim 0.5$, and converged into the optimal point, $\mu_0 \sim y_0$ and $\sigma_0\sim 0$. As we can confirm, they move almost along the vector field. However, we cannot get to the optimal one, $(\mu_0,\sigma_0)=(y_0,0)$, because that is numerically unstable. One of our network outputs, $\sigma_0$, is in a denominator of the loss function, equation (\ref{eq:Gloss}). In other words, our formulation of a variance network cannot have the optimal point as a solution for 1-point training. It is a reasonable result because we do not have a meaningful definition of standard deviation only with a data point.

\subsection{data density transition}
\begin{figure}
\begin{center}
\includegraphics[width=1.0\linewidth]{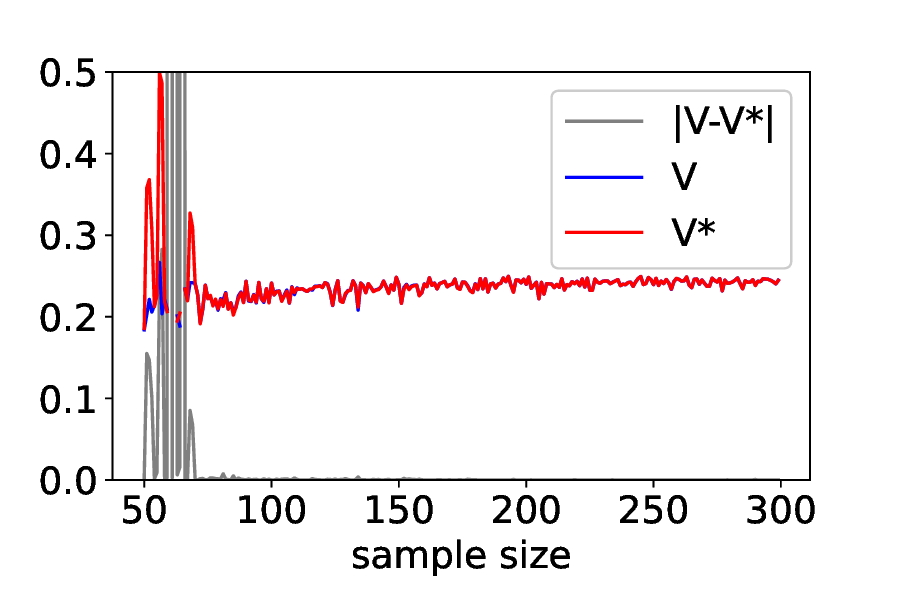}
\end{center}
\caption{Difference between two types of variance. We can estimate a variance with the mean squared error, $<(y_i-\mu_i)^2>$, plotted as 'V'. At the same time, we calculated another variance with the predicted standard deviation, $<\sigma_i^2>$, plotted as 'V*'. We also plot the difference between the two values, '|V-V*|'.}
\label{fig:density_transition}
\end{figure}

Next, we consider N-point training with a variance network. We sample pairs of random encoding, $(\bm{x}_i),y_i$, with a size, $N$, and train the network with it. Firstly we want to roughly grasp the feature of training dynamics. To do that we evaluate training results with variance. We can estimate it in two ways, mean squared error, $V\equiv<(y_i-\mu_i)^2>$, and another one, $V^*\equiv<\sigma_i^2>$. In FIG. \ref{fig:density_transition}, we show those estimated variances with different sizes of datasets, $N$. We can confirm the non-negligible difference between them in the cases with a little dataset, but it is negligible in the cases with a larger dataset. In FIG. \ref{fig:density_transition}, we show the difference after training of a fixed epoch, $e_{mx}=2000$, but it can depend on the duration of the training epoch itself.

\begin{figure}
\includegraphics[width=1.0\linewidth]{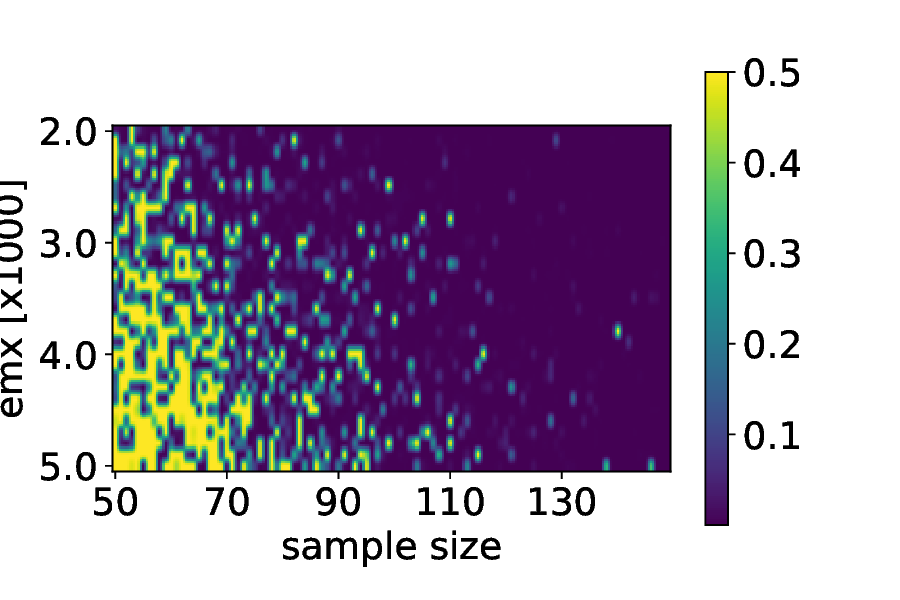}
\caption{Numerical instability against sample size and training epochs, $e_{mx}$. 
We show the difference between two predicted variances. The horizontal axis means the sample size of the training dataset. The vertical axis means training epochs. The difference is shown with color.}
\label{fig:density_transition2}
\end{figure}

In FIG. \ref{fig:density_transition2}, we show the dependence on the training epoch. In the figure, we plot the difference, $|V-V^*|$, against the size of a dataset, $N$, and the duration of the training epoch, $e_{mx}$. As we can see, the difference tends to grow when the duration, $e_{mx}$, is large. On the other hand, it tends to be reduced when the size, $N$, is large. In other words, the difference can be negligible when the data density is large but it may grow after enough training.

\begin{figure}
\begin{tabular}{cc}
\begin{minipage}{0.45\hsize}
\includegraphics[width=1.0\linewidth]{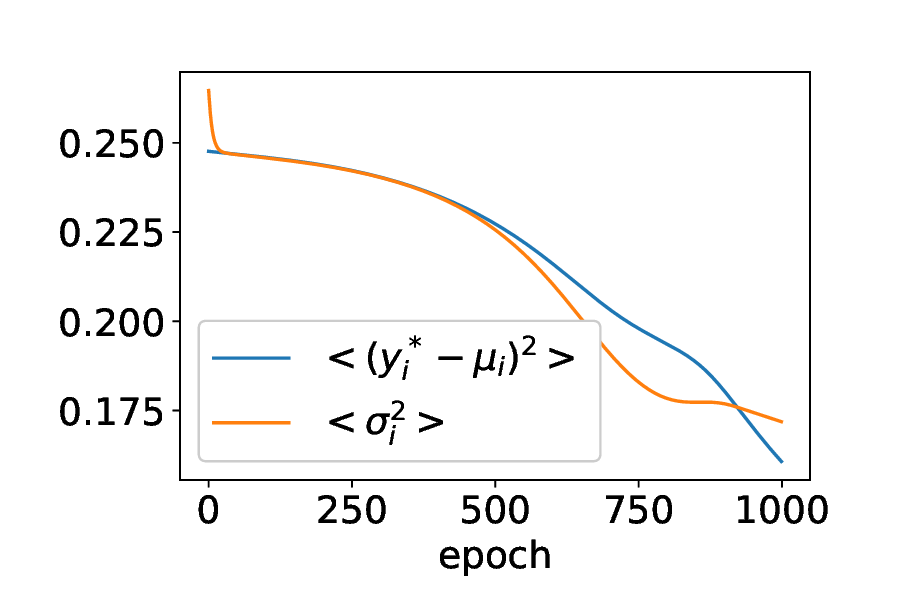}
\subcaption{Training dynamics of errors.}
\label{fig:var2}
\end{minipage}&
\begin{minipage}{0.45\hsize}
\includegraphics[width=1.0\linewidth]{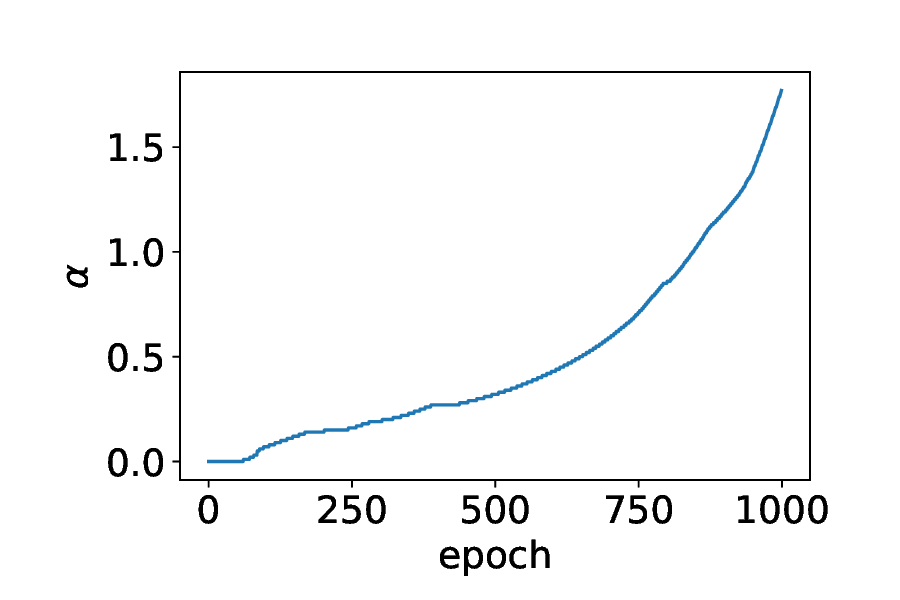}
\subcaption{Evolution of prediction scale.}
\label{fig:scale_alpha}
\end{minipage}
\end{tabular}
\caption{Training dynamics with a training dataset, $N=20$. (a) Two predicted errors, $<(y_i-\mu_i)^2>$ and $<\sigma_i^2>$, are shown. (b) We can approximate the predicted value, $\mu_i$, as a weighted average, $\sum_j \exp(-\alpha |\bm{x}_i-\bm{x}_j|)y_j/N$. Here we show the evolution of the prediction scale, $\alpha$.}
\label{fig:var-alpha}
\end{figure}

As we already confirmed, 1-point training is numerically unstable. {\color{black} In addition, N-point training is also unstable if the training dataset is sparse enough, FIG. \ref{fig:density_transition}.} If we assume the difference, $|V-V^*|$, stems from numerical instability, the results are reasonable. However, why the instability grows after longer training? We can expect that network outputs, $\mu_i$ and $\sigma_i$, depend not only on the local data point, $(\bm{x}_i,y_i)$, but also on other ones nearby it. We evaluate the output, $\mu_i$, with a weighted average, $\sum_j\exp(-\alpha|\bm{x}_i-\bm{x}_j|)y_j/N$. The parameter, $\alpha$, means a spatial scale of the average. In FIG. \ref{fig:var-alpha}, we show training dynamics with a dataset, $N=20$. On the left, we show the dynamics of two variances. On the right, we show the growth of the scale, $\alpha$. The scale is optimized so that the weighted average and the output, $\mu_i$, should match with each other. The variances show no difference at first. But we see a significant difference between them in the end. At the same time, the scale, $\alpha$, grows along the training. This suggests a spatial scale of prediction is reduced after enough training.
\subsection{kernel-balanced equation}
\begin{figure}
\begin{tabular}{cc}
\begin{minipage}[t]{0.45\hsize}
\includegraphics[width=1.0\linewidth]{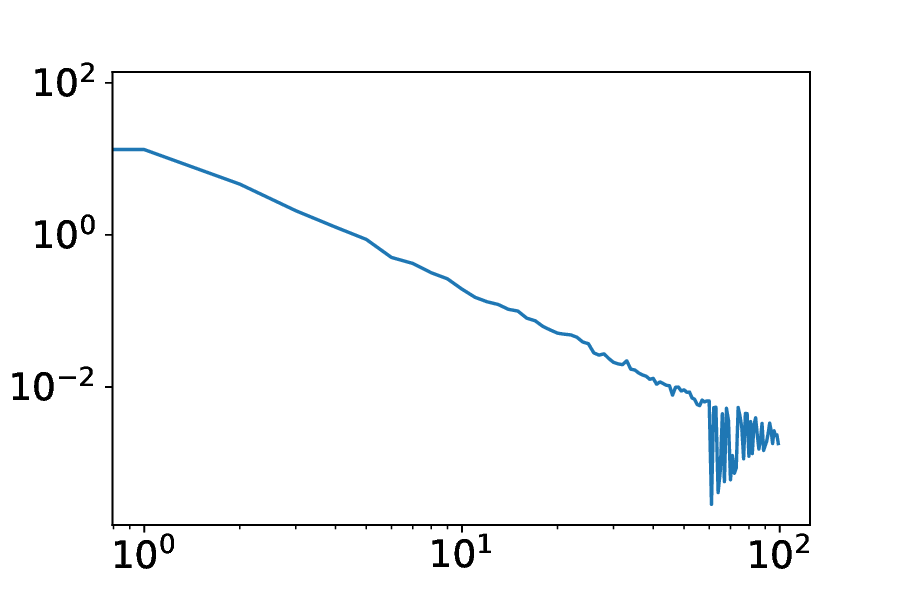}
\subcaption{Eigenvalues}
\label{fig:eigs}
\end{minipage}&
\begin{minipage}[t]{0.45\hsize}
\includegraphics[width=1.0\linewidth]{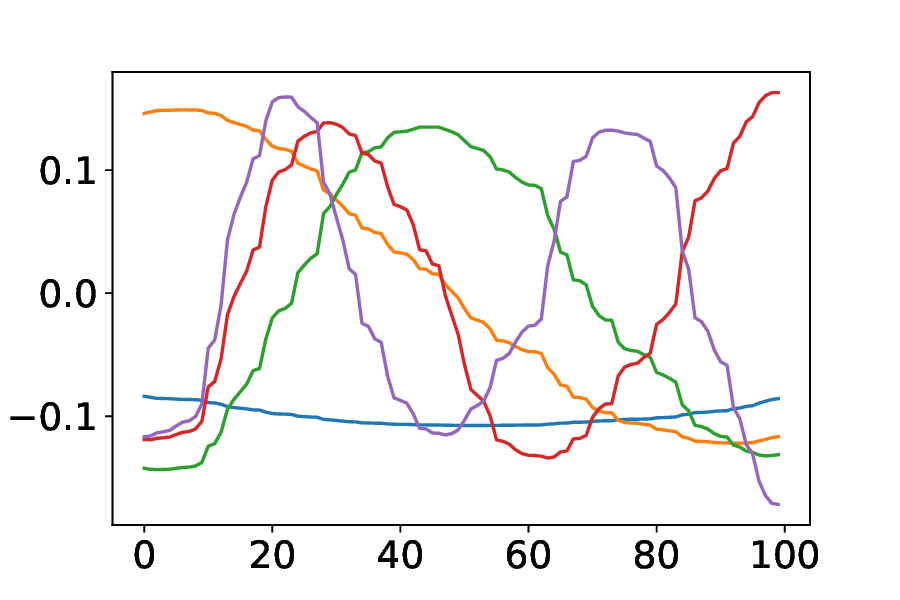}
\subcaption{Eigenvectors: 1st to 5th.}
\label{fig:eigv1}
\end{minipage}\\
\begin{minipage}[t]{0.45\hsize}
\includegraphics[width=1.0\linewidth]{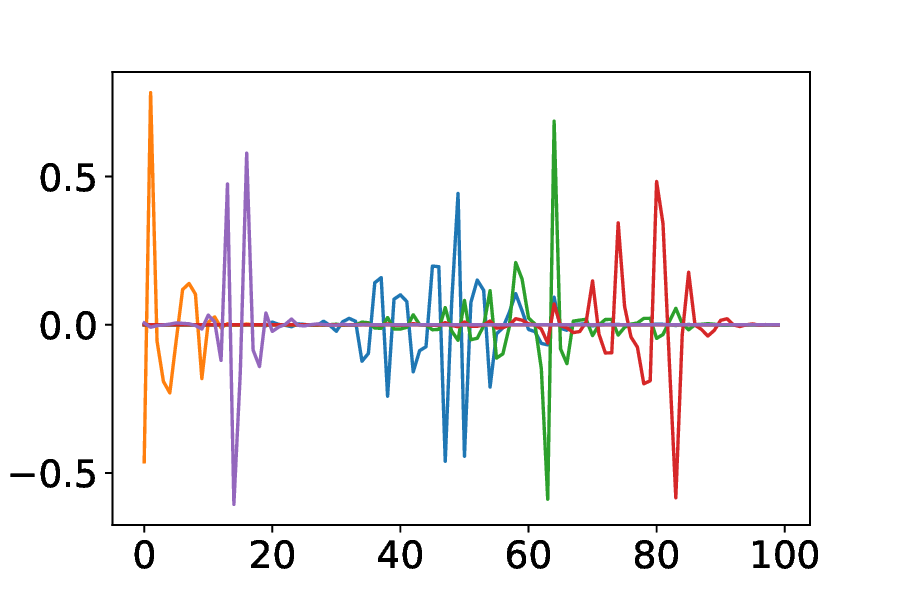}
\subcaption{Eigenvectors: 45th to 49th.}
\label{fig:eigv2}
\end{minipage}&
\begin{minipage}[t]{0.45\hsize}
\includegraphics[width=1.0\linewidth]{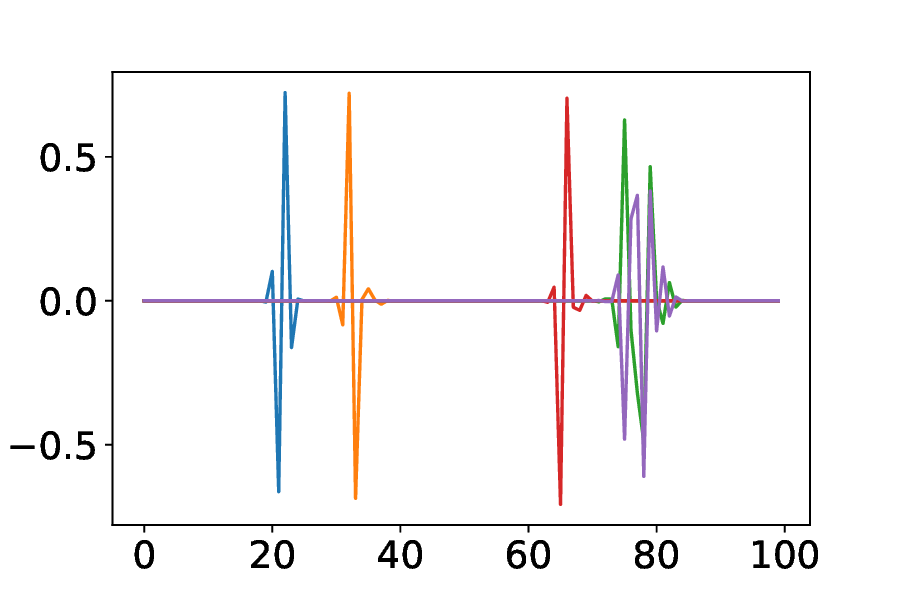}
\subcaption{Eigenvectors: 95th to 99th.}
\label{fig:eigv3}
\end{minipage}
\end{tabular}
\caption{Eigenvalues and vectors of a response kernel, $K_{ij}$. We calculated eigenvalues and vectors with a randomly distanced one within a range, $(0,1)$. We generated 100 random values, $x_i$, and made a response kernel, $\exp(-|x_j-x_i|)$, with those ones.}
\label{fig:K_ij}
\end{figure}

We consider the simplified dynamics, (\ref{eq:kernel_dyn}) or (\ref{eq:kernel_dyn2}), to understand the growth of scale, $\alpha$. Firstly, we study the training dynamics of an average network with its simplified one,
\begin{equation}
\frac{d\mu_i}{dt}\propto -\sum_jK_{ij}(y_j-\mu_j).
\end{equation}
The matrix, $K_{ij}$, consists of kernel distance terms, $K(|\bm{x}_i-\bm{x}_j|)$, and we assume it can be expressed in an exponential form as already introduced,
\begin{equation}
K_{ij}\sim\exp(-\beta|\bm{x}_i-\bm{x}_j|).
\end{equation}
We show eigenvectors and values with an ideally simple case, in FIG. \ref{fig:K_ij}. We constructed a matrix, $K_{ij}=\exp(-|x_i-x_j|)$, with sorted 100 random values, $0\leq x_i\leq 1$. In other words, we show the features of a response kernel with randomly distanced data points. As we can see, the eigenvalues decrease in a power-law manner. On the other hand, eigenvectors show Gabor wavelet-like forms\cite{wavelet,GW}. Major modes show broader waves than minor ones. This means that training dynamics reduces large scaled spatial error at first. Local error is reduced after enough training. These dynamics can be interpreted in our case as follows,
\begin{eqnarray}
\frac{d\mu_i}{dt}&\propto&\sum_jK_{ij}(y_j-\mu_j)\\
&\sim&\sum_jK_{ij}(y_j-\mu_i)\label{eq:Kaverage}\\
&\sim&\sum_k\lambda_k\sum_j\tilde{K}_{ijk}(y_j-\mu_i),
\label{eq:Kbalance}
\end{eqnarray}
where the values, $\tilde{K}_{ijk}\sim\exp(-\beta_k|x_i-x_j|)$, show differently scaled kernel terms. If we can assume a relation, $\mu_j\sim\mu_i\pm\delta$, around the point, $\bm{x}_i$, we get the equation, \ref{eq:Kaverage}. This means the local expectation, $\mu_i$, should match the weighted average, $K_{ij}y_j$. Finally, we rewrite it in a manner of multi-scale expansion, eq. \ref{eq:Kbalance}, using the differently scaled kernels, $\tilde{K}_{ijk}$, and the weight of each mode, $\lambda_k$. Needless to say, we assume the weight should be larger for the global averaging term, which has a small parameter, $\beta_k$.
Thus, the weighted balance of kernel response,
\begin{equation}
\mu_i=\frac{\sum_j\tilde{K}_{ijk}y_j}{\sum_j\tilde{K}_{ijk}},
\label{eq:KBequation}
\end{equation}
is realized from more global modes to local ones through training. 

In the same way, we can write down the kernel-balanced equation for a variance network in the following,
\begin{eqnarray}
\frac{d\mu_i}{dt}&\propto&\sum_j\frac{K_{ij}}{2\sigma_j^2}(y_j-\mu_j)\\
&\sim&\sum_k\lambda_k\sum_j\Tilde{K}_{ijk}\frac{y_j-\mu_i}{2\sigma_j^2}\label{eq:mudyn}\\
\frac{d\sigma_i}{dt}&\propto&\sum_j\frac{K_{ij}}{\sigma_j^3}((y_j-\mu_j)^2-\sigma_j^2)\label{eq:sigdyn1}\\
&\sim&\sum_k\lambda_k\sum_j\Tilde{K}_{ijk}\frac{(y_j-\mu_i)^2-\sigma_i^2}{\sigma_j^3}\label{eq:sigdyn2}.
\end{eqnarray}

We can notice the denominator, $\sigma_j$, in the equation (\ref{eq:mudyn}), as the difference from the previous one. In addition, we have one more equation on the other dynamics, in eq. (\ref{eq:sigdyn1}) and (\ref{eq:sigdyn2}). These equations suggest we can have numerical instability again, because the point, $\sigma_i=0$, is a local optimal as a result of training.

\begin{figure}
\includegraphics[width=1.0\linewidth]{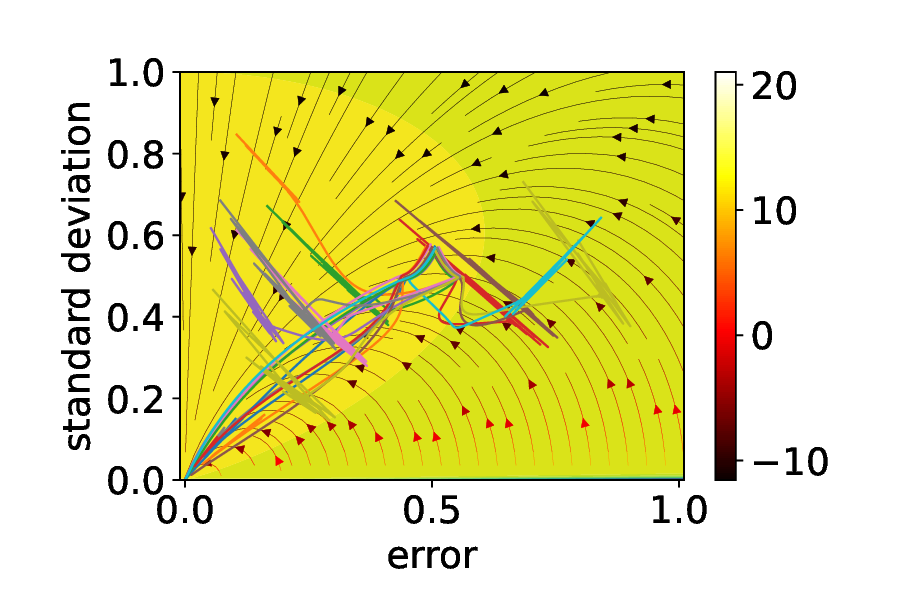}
\caption{Training dynamics with multiple training points, $N=20$. In the figure, we show all trajectories with different colors. The vector field shows the gradient of a loss function. The color of the arrows shows the steepness of the gradient in the log scale.}
\label{fig:point_training_N20}
\end{figure}

To confirm the numerically unstable dynamics, we show the training dynamics of N-point training, $N=20$, in FIG. \ref{fig:point_training_N20}. All the trajectories, $(|y_i-\mu_i|,\sigma_i)$, are plotted on the loss gradient, in the same manner as FIG. \ref{fig:point_training}. At a first glance, we notice jumps in them. Needless to say, these suggest numerical instability.

\begin{figure}
\begin{tabular}{cc}
\begin{minipage}[t]{0.45\hsize}
\includegraphics[width=1.0\linewidth]{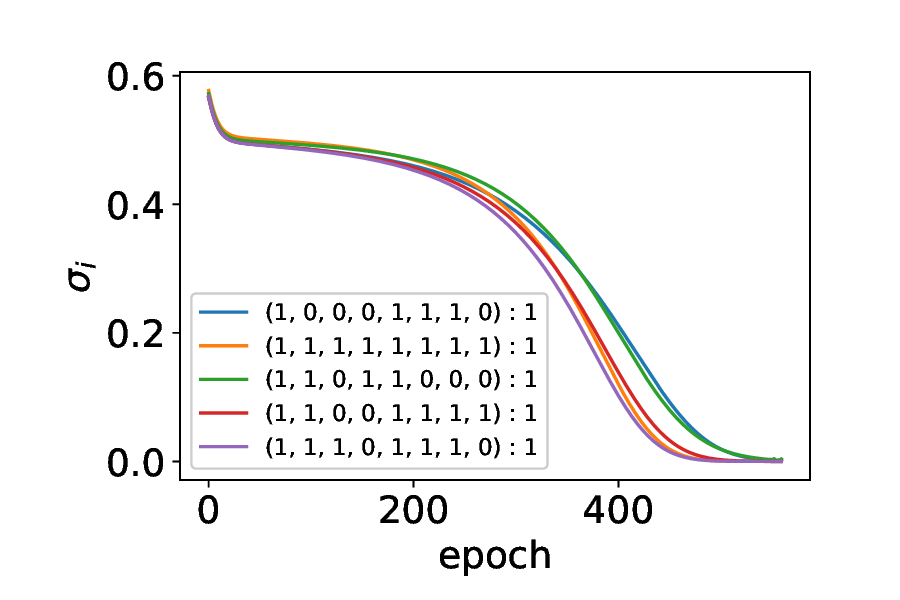}
\subcaption{Convergent group}
\label{fig:sigma_A}
\end{minipage}&
\begin{minipage}[t]{0.45\hsize}
\includegraphics[width=1.0\linewidth]{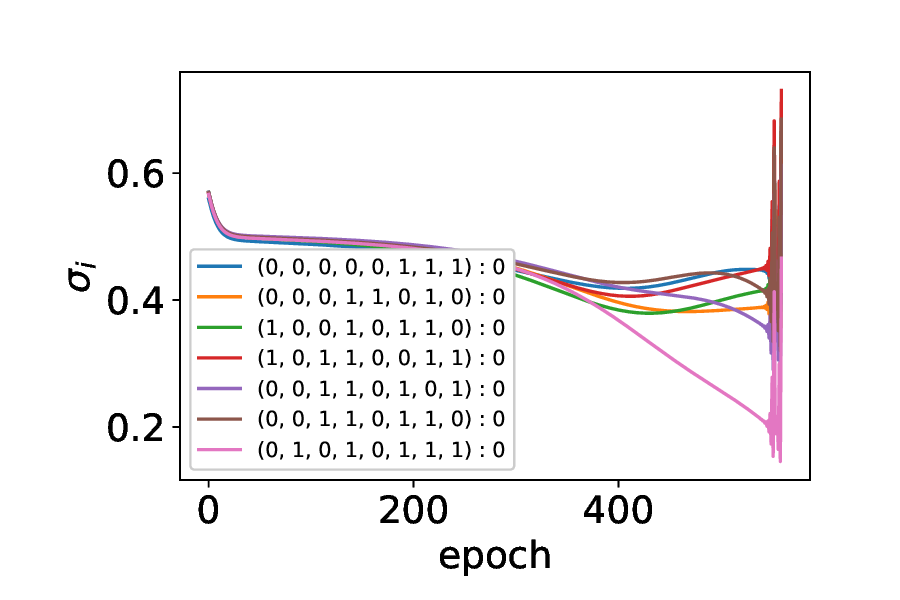}
\subcaption{Unstable group}
\label{fig:sigma_B}
\end{minipage}\\
\begin{minipage}[t]{0.45\hsize}
\includegraphics[width=1.0\linewidth]{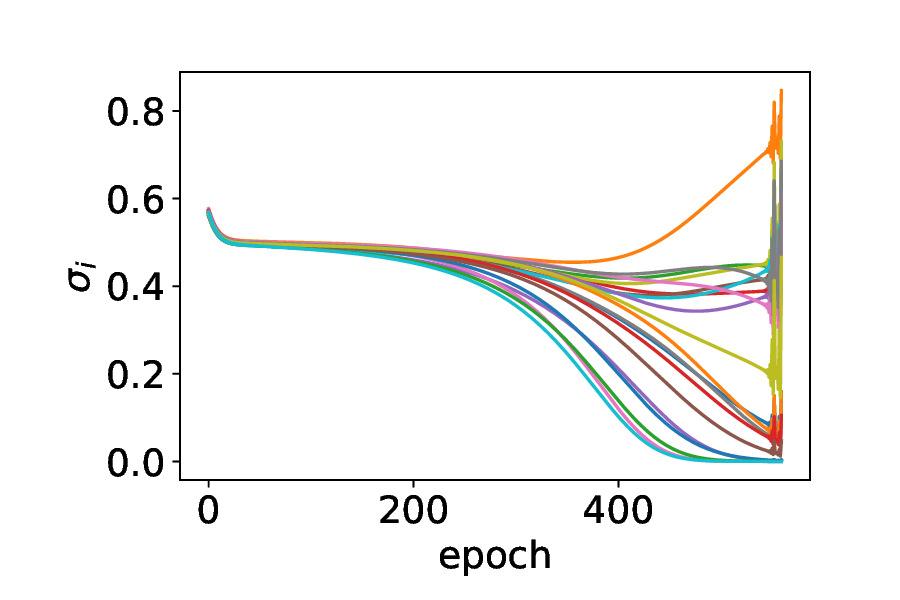}
\subcaption{Dynamics of $\sigma_i$}
\label{fig:sigma_Dyn}
\end{minipage}&
\begin{minipage}[t]{0.45\hsize}
\includegraphics[width=1.0\linewidth]{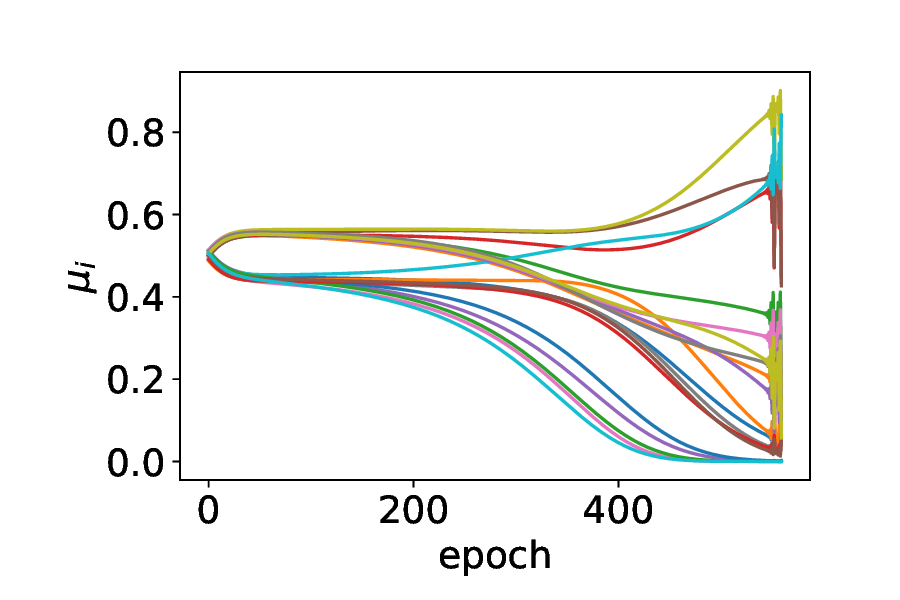}
\subcaption{Dynamics of $\mu_i$}
\label{fig:mu_Dyn}
\end{minipage}
\end{tabular}
\caption{A demonstration of numerical instability. We show training dynamics with sample size, $N=20$. (a) Trajectories of $\sigma_i$ for the group, converged into the state, $y_i^*-\mu_i=0,\sigma_i=0$. In the box, we show the pairs of input, $x_i$, and output, $y_i$. (b) Trajectories of $\sigma_i$, showing first instability. (c) All trajectories of $\sigma_i$. (d) All trajectories of $\mu_i$.}
\label{fig:N20_instability}
\end{figure}

We also show the training dynamics against the training epoch, in FIG. \ref{fig:N20_instability}. We show some convergent trajectories, in (a), and unstable ones, in (b). All of the trajectories, $\sigma_i$ and $\mu_i$, are plotted, in (c) and (d). We can confirm that jumps happen at the timing when the convergent trajectories approach the optimal.

\section{discussion}\label{sec:discussion}
Machine learning is now one of the most powerful ways to construct a model for a given dataset. Among the techniques in the field, deep neural networks play central roles not only in business applications but also in scientific studies. If we train a model with a dataset, $\{(\bm{x}_i,y_i)\}$, the trained model, $f$, can predict outputs for any input, $y=f(\bm{x})$, successfully even if the point, $\bm{x}$, is not included in the dataset. This feature is known as generalization. We know generalization should be a reflection of proximity in the input space. In other words, two outputs, $f(\bm{x}_a)$ and $f(\bm{x}_b)$, should be similar if those inputs, $\bm{x}_a$ and $\bm{x}_b$, are similar with each other. However, the distribution of the dataset is often very complicated and the spatial scale of similarity can be complicated as well. In this paper, the structure of the prediction scale is focused on. We consider neural tangent kernel or training response because they can describe the spatial effect of a training step and be applied to training dynamics. We elucidate the determination mechanism of the prediction scale with simple convolution networks and simplified models for the dynamics.

As an estimation problem for the dataset distribution, we adopt a loss function for the fitting to Gaussian one. We minimize the loss and get an optimal fitting model by training with a dataset. Our model, a variance network, outputs an expectation and standard deviation, $\mu(\bm{x})$ and $\sigma(\bm{x})$. As the dataset, we simply set a randomly generated encoding from a 1D bit-string, $\bm{x}_i$, to a binary output, $y_i$.
This problem set is a very simple one so that we can understand the dynamics of the prediction scale.

As shown in FIG. \ref{fig:point_training}, the dynamics is numerically unstable in the simplest case, 1-point training, because we cannot determine a meaningful standard deviation. In a similar manner, N-point training suffers from numerical instability, as shown in FIG. \ref{fig:density_transition} and FIG. \ref{fig:density_transition2}. The numerical instability stems from the reduction of the prediction scale, shown in FIG. \ref{fig:var-alpha}. The scale gradually reduces along training and finally two types of variances, $<(y_i-\mu_i)^2>$ and $<\sigma_i^2>$, do not show consistency with each other.

To understand the dynamics of the prediction scale, response kernel dynamics for an average network are studied. The kernel matrix has Gabor wavelet-like eigenvectors with eigenvalues decaying along a power law. This suggests the network learns spatially larger scale patterns at first and local patterns later. These dynamics can be shown in a kernel-balanced equation, in eq. (\ref{eq:KBequation}). An average network outputs a weighted mean value averaged over the answers, $y_i$. The weight determines the prediction scale and it decreases along the training. In the case of variance network, the solution is not so straightforward, eq. (\ref{eq:sigdyn1}) and (\ref{eq:sigdyn2}), but the prediction scale decreases along training as well. In addition, the predictions suffer from numerical instability again. Once any prediction, $(\mu_i,\sigma_i)$, approaches to the optimal point, $\sigma_i=0$, it destabilizes training dynamics of other predictions, $(\mu_j,\sigma_j)$.

The kernel-balanced equation suggests we can understand convolutional networks as a function that outputs the local average. The scale of averaging is not fixed and decreases along the training. In other words, the network can output the same values as the dataset after enough training, if the response kernel does not collapse during it. It is known that the kernel is constant in an ideal condition. Even if the network is not ideal and has a finite size, it can be seen as almost constant in some cases. We still need more studies on kernel stability but we believe the equation should be effective for a more variety of cases.

The problem we consider here is known as uncertainty estimation in the field of machine learning. In reality, we suffer from uncertainty for many reasons in practical usage. If we should assume some noises in the observation, we cannot regard the dataset as the ground truth anymore. Even if we can exclude such a noise in a way, the truths may not be deterministic. In addition, we have some more origins of it in the model and its training. The model can be redundant and have many solutions for the given dataset. We often use non-deterministic training algorithms and this can result in many solutions.

It is known that uncertainty can be divided into two classes, epistemic and aleatoric one\cite{uncertainties}.
What we consider here is the latter one. In such a case, we can model the dataset as a distribution at most. In this context, what we show here is the insufficiency of Gaussian modeling. Our formulation suggests Laplacian modeling is insufficient as well. On the contrary, a Bayesian approach, in which the parameter has its distribution, can be a solution. In fact, the numerical instability occurs at the optimal point, $\sigma_i=0$, and can be overcome by replacement of it, $\sigma_i$, with another probabilistic one, $P(\sigma_i)$. Actually, it is reported that t-distribution is an effective way of that\cite{t-variance}. Our formulation explains the reason for the effectiveness, therefore.

{\color{black} In our days, we often train very large models with large datasets. In such a case, the dataset usually distributes in a non-uniform manner. As pointed out, we often see model variability in such cases\cite{mod-var}. As we have shown with the kernel-balanced equation, the prediction can be different between training phases, especially for minor modes in the dataset. Since the training for minor modes is postponed to a much later stage depending on the initial condition, we can see model variability.}


{\color{black} Our instability stems from the shape of the loss function, equation \ref{eq:Gloss}. As shown in the kernel-balanced equation, this sort of instability cannot be prevented in a straightforward way. However, when the model and the dataset are very large, training requires much longer epochs and we often need much larger computing time for a training epoch. In such a case, we may not see instability because of short training time, in reality. As a possibility, when we have drop-out layers in the network, we can have a non-zero variance in the prediction for the same input. If the standard deviation for the input, $\bm{x}$, is not zero, we do not see the instability. In this meaning, the instability is not necessarily universal for real applications. However, the analysis with the kernel-balanced equation, especially for the average network, can be applicable to a wide range of applications, because of its simplicity. As the kernel-equation equations shows, training always starts from major modes and proceeds to minor modes later, we can carefully design the dataset density for more efficient training, though we still need more studies.
}


As such an application, we can apply it for variance estimation without a variance head for prediction of standard deviation, $\sigma(\bm{x})$. Since the output after $t$-epochs, $f_t(\bm{x})$, approximates a local average dependent on the prediction scale, the difference between two outputs, $(f_t(\bm{x})-f_T(\bm{x}))^2$, tells us the local variance in the limit, $T\rightarrow\infty$. We do not know the quantitative precision of this formula yet, but it can be a convenient way for variance estimation.

We are getting more and more data through ubiquitous sensors and it is even available on the web. The field of data science sheds light on the complexity of the real world with them. Actually, astonishingly powerful applications emerge one after another with the aid of such huge datasets and machine powers\cite{med_image,med_vision,chatgpt4,auto_driving}. Machine learning algorithms are necessary not only in such practical applications but also in scientific studies in the real world's complexity\cite{DM1,DM2,FRG}. However, we note that those algorithms still require further understanding.
In reality, new technologies often focus on an innovative mathematical formulation and its implementation, but a dynamical understanding must be necessary as well\cite{TR,gan_dyn}. We can design an efficient and even safe one with such an understanding. We believe the science of complexity can be an effective approach for the field.
\section*{Acknowledgements} \label{sec:acknowledgements}
This work is motivated through discussions in SenseTime Japan, HONDA, and an internship student WL.


\end{document}